# Subtle pH differences trigger single residue motions for moderating conformations of calmodulin

**Ali Rana Atilgan, Ayse Ozlem Aykut, and Canan Atilgan\***

Faculty of Engineering and Natural Sciences, Sabanci University, 34956 Istanbul, Turkey

\*Corresponding author

*e-mail*: canan@sabanciuniv.edu

*telephone*: +90 (216) 4839523

*telefax*: +90 (216) 4839550

**ABSTRACT**

This study reveals the essence of ligand recognition mechanisms by which calmodulin (CaM) controls a variety of $Ca^{2+}$ signaling processes. We study eight forms of calcium-loaded CaM each with distinct conformational states. Reducing the structure to two degrees of freedom conveniently describes main features of conformational changes of CaM via simultaneous twist-bend motions of the two lobes. We utilize perturbation-response scanning (PRS) technique, coupled with molecular dynamics simulations to analyze conformational preferences of calcium-loaded CaM, initially in extended form. PRS is comprised of sequential application of directed forces on residues followed by recording the resulting coordinates. We show that manipulation of a single residue, E31 located in one of the EF hand motifs, reproduces structural changes to compact forms, and the flexible linker acts as a transducer of binding information to distant parts of the protein. Independently, using four different pKa calculation strategies, we find E31 to be the charged residue (out of 52), whose ionization state is most sensitive to subtle pH variations in the physiological range. It is proposed that at relatively low pH, CaM structure is less flexible. By gaining charged states at specific sites at a pH value around 7, local conformational changes in the protein will lead to shifts in the energy landscape, paving the way to other conformational states. These findings are in accordance with FRET measured shifts in conformational distributions towards more compact forms with decreased pH. They also corroborate mutational studies and proteolysis results which point to the significant role of E31 in CaM dynamics.

*Abbreviations.* CaM, calmodulin; PRS, perturbation-response scanning; $Ca^{2+}$-CaM, $Ca^{2+}$ loaded CaM; MD, molecular dynamics; CS, conformational space

**INTRODUCTION**

The functional diversity of proteins is intrinsically related to their ability to change conformations. As a notorious example, calmodulin (CaM) has the pivotal role of an intracellular $Ca^{2+}$ receptor that is involved in calcium signaling pathways in eukaryotic cells [1]. CaM can bind to a variety of proteins or small organic compounds, and can mediate different physiological processes by activating various enzymes [2,3]. Binding of $Ca^{2+}$ and proteins or small organic molecules to CaM induces large conformational changes that are distinct to each interacting partner [1,3,4]. In fact, the interaction of CaM with target proteins at various levels of $Ca^{+2}$ loading control many key cell processes as diverse as gene expression, neurotransmission, ion transport; see [5] and references cited therein. Also, diseases related to unregulated cell growth, such as cancer, have been shown to have elevated levels of $Ca^{2+}$ loaded CaM ($Ca^{2+}$-CaM) [6].

Structural heterogeneity of CaM depends significantly on the environmental conditions of pH, ionic strength, and temperature. The two opposing domains act uncoupled at low $Ca^{2+}$ loading, while stabilization of especially the C-terminal domain upon $Ca^{2+}$ loading leads to coupled motions between the domains, with concerted rotational dynamics occurring on the order of 15 ns time scale [7]. The coupling is orchestrated by the flexible linker region, letting the two domains adopt a large distribution of relative orientations so that its wide variety of different targets may be accommodated [8,9]. The coupling between the domains is lost at higher temperatures or acidic pH [8,10,11] possibly due to the increased flexibility of the linker [12]. Direct measurement of the conformational distributions is now made possible via single molecule experiments [13], as well as combined ion mobility-mass spectrometry methods [14], disclosing at least three distinct regimes adopted by CaM structures. NMR experiments also point out that $Ca^{2+}$-CaM adopts a distribution of conformations [15], whereby neither the originally observed dumbbell shaped [16] nor the later recorded much compact crystal structures [2] are in abundance in solution. Such conformational plasticity of $Ca^{2+}$-CaM was further demonstrated by disorder analysis of crystallographic data [17]. Single molecule experiments have also established the distribution of possible structures and how they shift with change in environmental conditions such as $Ca^{2+}$ concentration, pH, and/or ionic strength [18]. Furthermore, macromolecular crowding was shown to stabilize the more collapsed conformations [19]. Recent fluorescence correlation spectroscopy experiments have quantified the time scale of interconversions between the various available states to be on the order of 100 μs [20].

In this manuscript, we aim to provide an explanation for the significantly large conformational changes occurring in response to small perturbations that may arrive at local regions of $Ca^{2+}$-CaM. It is known that comparison of experimentally determined ligand bound/unbound forms of a protein gives a wealth of information on the basic motions involved, as well as the residues participating in functionality [21,22,23,24]. Due to the time scales involved (100 μs [20]) that are much slower than may be observed by molecular dynamics (MD) simulations (100 ns), other computational and theoretical approaches must be employed. To decipher the key residues that may be targeted to facilitate ligand binding to $Ca^{2+}$-CaM, we employ the perturbation-response scanning (PRS) technique, coupled to MD simulations [21].

In the literature, a variety of computational techniques is available to get information from the Protein Data Bank (PDB) structures [25]. Most of these analyses reveal the different modes that may be stimulated by various ligands binding to the same apo form [24,26,27,28,29,30,31]. However, there is lack of information on how these modes are used by different ligands acting on the same protein. It is also unclear how many modes are stimulated by the binding of ligands. In particular, if the conformational change involved is more complicated than, e.g. hinge bending type motion of domains, several modes may be operational at the same time to recover the motion observed. Studies showed that collectivity is detrimental to the ability of representing the motion by a few slow modes [32,33]. We have recently put forth a new measure that we term redundancy index (RI) to quantify such collectivity [22]. RI is identified as the ratio of two contributing effects, (i) the local clustering around the residues that survives under loss of instantaneous interaction between pairs of atoms, and (ii) the global efficiency of communication of a given residue with the rest of the protein.

To get useful information similar to those obtained in experiments, it is of utmost interest to use a methodology that puts the system slightly out of equilibrium, and monitors the evolution of the response. Experimentally, the perturbation given to the system may be in the form of changing the environmental factors (e.g. changes in ionic concentration [34]), or may target specific locations on the structure itself, either through chemically modifying the residues (inserting mutations) [35] or by inducing site-specific perturbations (e.g. as is done in some single molecule experiments [36], or through ligand binding). Theoretically, the perturbation might be a force given to the system mimicking the mentioned forms above. The response of the system is then recorded to detect the underlying features contributing to the observations, yielding additional information than the operating modes of motion discovered.

It is possible to find examples of studies comprising of such approaches in the literature, all operating in the linear response regime. In one all-atom study, the perturbation has been applied as a frozen displacement to selected atoms of the protein, following energy minimization, the response has been measured as the accompanying displacements of all other atoms of the protein [37,38]. This has led to finding the shifts in the energy landscape that accompany binding [37,39]. This method based on molecular mechanics, scans all of the residues to produce comparative results. In various studies, the perturbations on residues are introduced by modifying the effective force constants [40], links between contacting residue pairs [31,41], or both [42]. On the other hand, perturbations may be inserted on the nodes instead of the links between pairs of nodes; depending on the location of the perturbation, the resulting displacements between the apo and holo forms may be highly correlated with those determined experimentally [43]. The more recent PRS methodology has successfully demonstrated that the conformations of a variety of proteins may be manipulated by single residue perturbations [22]. Using PRS to study in detail the ferric binding protein (FBP), it has been possible to map those residues that are structurally amenable to inducing the necessary conformational change [21].

In this work, we study the rich conformational space of $Ca^{2+}$-CaM by applying PRS to its unliganded extended form. We strive to understand conformational motions utilized to achieve the ligand bound forms of six different CaM structures, where the size of the ligands varies from a few atoms to peptides of 26 residues long. We also study the interconversion to the compact unliganded form of $Ca^{2+}$-CaM.

The manuscript is organized as follows: In the Results section, we first analyze these eight structures, to discern the types of conformational changes involved. We then describe the crude results obtained via PRS, and we compare the findings with predictions from slow modes of motion of the protein. Next, we reduce the protein to a highly simplified system with a few degrees of freedom, and demonstrate how single residue perturbations may lead to large changes in this highly coarse-grained picture of the protein. We also identify the relation between the single residues/directions found to participate in the conformational change and allosteric communication in CaM. All results are examined in conjunction with experimental observations reported in literature. In the Discussion section, we summarize how changes in the electrostatic environment of the protein, quantified by shifted pKa values of charged residues, may lead to generating mechanical response; we put forth PRS as a robust technique to uncover such cause-effect mechanisms. In the Materials and Methods section, we describe the structures studied, the PRS technique and the molecular dynamics (MD) simulations we use to generate the variance-covariance matrix which is the operator in PRS. We also summarize the pKa calculation methods used in this study.

**RESULTS**

**A survey of the protein structures.** In this study, we explore the conformational change in $Ca^{2+}$-CaM upon binding six different ligands. As listed in Table 1, these ligands have as many as 26 residues, and they bind to various regions of the protein. A variety of conformational changes are observed upon binding to the fully calcium loaded CaM ($Ca^{2+}$-CaM) (figure 1). In addition, we study the conformational jump between the extended and compact forms of unliganded $Ca^{2+}$-CaM (1prw).

For every pair in the set, we perform STAMP structural alignment, implemented in VMD 1.8.7 MultiSeq plugin [44]. We record the root mean square deviation (RMSD) between the structures of the target forms with the extended, initial structure (gray shaded column in Table 2). We also record separately the RMSD of the N-lobe and C-lobes (Table 2). We find that the overall RMSD between the initial and target structures are mostly on the order of $15 - 16$ Å, except for 1rfj (2.7 Å) and 1mux (6.4 Å). We note, however, that the magnitude of the change does not depend on the ligand size, or the region of the protein it binds. In contrast, the superposition of only N- or C-domains yields low RMSD (gray shaded row in Table 2); the only ones that have values above 1.5 Å are the N-lobes of 1prw (2.3 Å) and 2bbm (1.9 Å). Thus, the internal arrangements in the two lobes are nearly the same. This hints that the conformational change mostly involves global motions rather than local rearrangements.

We also compare in Table 2 the RMSD amongst the seven target structures themselves to quantify the amount of structural difference they have. 1rfj and 1mux both have 14 – 16 Å RMSD with the other five structures. Since these two are also the ones that have smaller RMSD with 3cln, we conjecture that they may be located closer to the extended form in the conformational space (CS), and at a different part than the other five target structures. They are not, however, in exactly the same region of the CS since the overall RMSD between them is large (7.4 Å). The internal arrangement of both the N- and the C-lobes are similar with 1.3 and 1.1 Å RMSD, respectively, so the structural difference must be in their relative positioning. Also displayed in Table 2 are the overlaps between the displacement vectors of the experimental structures (equation 5). Inspecting the $O^{jk}$ values listed in Table 2, we find that 1lin, 1prw, 1qiw, 2bbm and 1cdl display similar types of conformational motions ($O^{jk} \geq 0.89$), while 1rfj and 1mux each have distinct conformational changes from the rest ($O^{jk}$ is in the range 0.03 – 0.28), as well as from each other ($O^{1rfj-1mux} = 0.20$). Taken together, the RMSD and overlap values imply a closing of the two lobes towards each other. Thus, despite the variety of ligand types and ligand sizes in the bound forms, there are three classes of conformational changes. 1lin, 1qiw, 2bbm and 1cdl are stabilized in the closed conformation of $Ca^{2+}$-CaM, exemplified by 1prw. This is in addition to the unique forms of 1rfj and 1mux.

The most similar pair of structures is 1qiw/1lin, which both have large groups binding in the region between the two lobes. Moreover, the internal structures of the N- and C-lobes are almost the same [(RMSD is 0.7 Å for both lobes, much below the resolution of the x-ray experiments (Table 1)]. The remaining pairs of target structures have RMSD in the range of 1.9 – 5.2 Å. In some cases, the N-lobe RMSD may be well above the experimental resolution, e.g. as high as 3.3 Å for the pair 1prw/2bbm. This is in contrast to the rigidity of the C-lobe, which has an RMSD of less than 1.5 Å in all cases. The observation of considerably less mobility in the latter lobe is in accord with the higher affinity of the C-domain for $Ca^{+2}$ (at 7 μM $Ca^{+2}$ concentration) at as opposed to the N-domain (at 300 μM) [45].

**Directionality matters for conformational change preferences of CaM.** By using PRS, we sequentially insert random forces on each residue. For each residue, $i$, we then compute the overlap coefficient, $O^i$ (equation 3) between the response vector $\Delta \mathbf{R}^i$ and the experimental conformational change vector $\Delta \mathbf{S}$. In Table 3, we report the results of the PRS analysis for the seven target structures, using the variance-covariance matrix of the last 90 ns portion of the trajectory. The reported values represent both the single best overlap obtained, as well as the averages over 500 independent scans. We observe that for none of the conformational changes it is possible to obtain a high overlap by perturbing a single residue in a randomly chosen direction; the best average random perturbation on a single residue yields a value of 0.56. This is in sharp contrast to our earlier study on FBP [21] where perturbation of an allosteric site is independent of the directionality of the perturbation. However, in a later study where we conducted PRS on 25 proteins, we found such a simplistic result only for a subset of proteins, most of which were comprised of those displaying hinge motions of the domains [22].

For CaM, we do find that it is possible to mimic the different conformational changes to an overlap of 0.70±0.03 for five of the target forms by acting on residue E31 in a selected direction. Neighboring residues to E31 also yield comparable overlap in some cases; these reside on the edge of one of the EF-hand motif loop I (see **Proteins** subsection under **Materials and**

**Methods**). For the case of 1mux, perturbing the $Ca^{+2}$ ion residing in loop I significantly improves the overlap to 0.43, although this value is well below those for the other target structures. Thus, the initial extended CaM structure may be manipulated from this particular EF hand motif.

We also find that L69 appears in manipulating 3CLN towards three of the target structures (Table 3). To identify if coupled conformational manipulation improves the results, we have perturbed residues 31 and 69 in pairs. 5000 random perturbations inserted simultaneously on 31/69 did not lead to any improvement of the overlaps. Finally, we have made 500 independent scans of coupled perturbations of E31 with all other residues, and did not identify any improvement of the overlaps. Moreover, we have made a 100 iteration scan of all possible node pairs (i.e., $100 \times 147^2 = 2160900$ independent pair force insertions) and confirmed that single node perturbations lead to the maximum overlap results for 3cln.

**Can conformational changes of CaM be described by slow modes of motion?** By inspecting Table 2, we have already made the observation that, some of the target structures may be represented by the same displacement vector $\Delta \mathbf{S}$. They may also have low RMSD within the lobes so that the overall conformational change is represented by the relative positioning of the two lobes. This might imply that the motion is described by a single dominant mode, which we will now show is not the case.

We seek the mode that best represents the conformational change by calculating the overlap of each eigenvector of the variance-covariance matrix and the experimental conformational change vector between the 3cln and target structures. We find that four modes dominate the largest overlaps. However, depending on the MD simulation chunk we are investigating, these may be any one of the most collective four modes. In other words, the precedence of the eigenvector changes between the different chunks of MD simulations, while its shape remains the same (e.g. eigenvector 1 calculated from the 1–10 ns interval of the simulation overlaps with eigenvector 2 calculated from the 10 – 20 ns portion ($O = 0.86$); conversely, eigenvector 2 of the former overlaps with eigenvector 1 of the latter ($O = 0.89$).

We therefore identify the lowest four modes as describing the following motions. We emphasize that the numbering is arbitrary since these modes appear in changing orders in different portions of the trajectories. The highest overlaps are listed in Table 3 along with the mode number:

Modes I and II both represent bending of the two lobes towards each other, while the planes in which the bend occurs are orthogonal. Mode I appear as partly describing the conformational change of five target structures with overlaps in the range 0.45 – 0.49. On the other hand, the conformational change of 1rfj is represented by mode II with an overlap of 0.52, while this mode is not representative of any of the other conformational changes. Mode III may be best described as a second bending motion. It partially represents the conformational change of 1mux with an overlap of 0.30. In many of these target forms, modes I and III both partially describe the conformational motion; yet, the two of them together do not improve the prediction in 2bbm, 1cdl and 1mux while they partially improve that of 1lin and 1prw. Mode IV corresponds to a rotation of the two lobes around the extended linker axis, while the linker remains almost rigid. None of the observed conformational changes are represented by such a motion.

In sum, the conformational changes of five of the seven target forms studied here are best recovered by a force applied on a single residue with overlap coefficients of ca. 0.7. The modal analysis shows that, there is no single low frequency (collective) mode, nor few multiple modes, that best describes these changes. Thus, the perturbation of a single residue (E31) must be invoking multiple modes of motion in these structures which shift from open to closed conformations. The conformational change of 1rfj is recovered well by manipulating the structure from its N-terminus (the first four residues are missing in the 3cln structure) which predominantly induce the single collective mode II. This is a bending motion that may also be observed by visually inspecting the two structures (figure 1). Finally, for 1mux, inspection of figure 1 shows that the main change is due to the destroyed linker conformation, since the N-lobe and C-lobe conformations are relatively intact (see the intra-lobe RMSD values in Table 1). Such flexibility-caused conformational changes may be recovered neither via dominant modes nor via single residue perturbations.

**A twist and a bend overcome a local free energy barrier.** We may group the target protein structures according to the findings until this point: Group 1 consists of 1lin, 1prw, 1qiw, 2bbm, and 1cdl where the change is best captured by perturbing E31 and its immediate neighbors. Group 2 has 1rfj whose motion is described by simple bending. Group 3 has 1mux whose conformational change is only partially described by either a perturbation or a collective mode.

To better understand the conformational motions CaM is capable of, we reduce it to three units, made up of the N-lobe, the flexible linker and the C-lobe. From the RMSD values in Table 2, we know that the internal atomic rearrangements in the C-lobe is almost non-existent and those within the N-lobe is relatively low (in the range of $0.4 - 3.3$ Å). In contrast, the RMSD between the structures may be as high as 16 Å, which must mainly be coordinated by the flexible linker. This viewpoint is supported by NMR results whereby multiple conformations of $Ca^{+2}$-CaM were discussed from the perspective of the linker [15]. That model yields compatible solutions to the experimentally measured nuclear coordinate shifts and residual dipolar couplings if the linker is modeled flexibly in the range of residues 75-81, while the N- and C-terminal domains are assumed to be rigid. The analysis also suggests that all sterically nonhindered relative conformations of the two domains are not equally probable, and that certain conformations are preferred over others in solution.

Thus, we use a simplified set of coordinates to capture the main features of the relative motions of these units by reducing the structure to five points in space. These points are schematically shown in figure 2a. Three of these points are the center of masses (COMs) of the N terminus (point 1), C terminus (point 5) and the linker (point 3). In addition, residues 69 and 91 are used to mark the beginning and end points of the linker (points 2 and 4). We then define two main degrees of freedom to capture the essence of the motions of the two lobes relative to each other. Angle $\theta$ defines the bending motion observed between the N- and C lobes, while $\varphi$ defines the relative rotation of the two lobes around the linker as a virtual dihedral angle (figure 2a). We note that similar virtual dihedral angle definitions on CaM were previously made [46,47,48,49]. It is also possible to define angles $\theta_N$ and $\theta_C$ for the bending motion observed between a given terminus and the linker (also shown in figure 2a). However, we found these degrees of freedom are not descriptive of the conformational change (for all proteins studied $\theta_N = 140 \pm 30°$ and $\theta_C = 100 \pm 30°$ without any distinguishing feature), and we do not discuss them further.

The joint probability distribution of the $\theta$ and $\varphi$ value pairs computed throughout the 120 ns long MD trajectory for the $Ca^{+2}$-CaM are shown as a contour plot in figure 2b along with the location of the eight PDB structures studied in this work. This verifies that the conformations visited during the trajectory are sampled around one minimum. We find that the region sampled in MD only visits one of the target structures, 1rfj whose motion is described by simple bending. The bending angle of the linker changes by $\pm 20°$ throughout this time window, and it essentially maintains the collinear arrangement of the two lobes ($<\theta> = 162°$). The torsional motion of the two lobes with respect to each other is more variable, $\varphi$ changing by $\pm 50°$ in this time window.

The average value of the torsion $<\varphi>$ is 67° so that the main axes of the two lobes are nearly at right angles to each other in space.

The initial structure, shown by the empty circle, is often visited during the MD sampling, while only one of the the target structures (1rfj) is reached within the 120 ns time window. Group 1 proteins are all located in one part of the reduced conformational space, shifted to higher $\varphi$ and lower $\theta$ values; i.e. they quantify the compact forms we observe in figure 1. In this reduced CS, it is also evident that 1rfj and 1mux are located closer to the initial structure, each occupying a unique part of the CS. These two structures are different from each other, though, the former having a more extended form. We also know that they do not occupy the same extreme states flanking the free energy minimum of the apo form. If this were the case, the conformational change of both of these structures would have been described by the same slow mode (Table 3).

PRS captures the conformations which are located far apart in the coarse grained conformational space, by giving perturbations to the same single residue. These perturbations are direction specific. In figure 3a we present the perturbation (red thick arrow) and the response (green arrows) that leads to the maximum overlap for 1lin. We observe that the response is a simultaneous bending of the two lobes towards each other accompanied by the twisting of the linker. Such twisting motions are of much higher frequency compared to the most collective ones. For example, moderate modes 6-20 have such twists, but each carry a partial motion of the linker as opposed to the overall twisting of the whole linker shown in the figure. Thus, the diagnosis of the PRS method is that the main motion governing the conformational change is a collection of the slow and moderate modes, and that they may be best described as a twist and a bend for the group 1 molecules. It also demonstrates that it is possible to simultaneously induce them via a single residue perturbation with the correct directionality. These observations are supported by a Monte Carlo study on helix models that suggests applied torques along with constraints on the ends of $\alpha$ helical regions lead to a nonlinear coupling between the bending and extensional compliances [50].

**E31 is a signaling residue for global communication in CaM through the linker.** PRS analysis reveals that E31 on the N-lobe (along with its immediate neighbors) consistently emerges as an important residue in manipulating the extended $Ca^{+2}$-CaM structure towards the compact conformations of many of the observed target structures. In this subsection, we shall further concentrate on E31. Our analysis is based on structural considerations, but there is plethora of previous work on CaM which implicates this residue occupies an important location affecting the dynamics in apo CaM as well as partially or fully $Ca^{+2}$ loaded CaM.

For example, E31 was implied to be involved in interdomain interactions of $Ca^{+2}$-CaM in an EndoGluC footprinting study [51]. EndoGluC proteolysis specifically cleaves at non-repeating glutamate sites of which there are 16 in CaM. The results point to E31 as a unique site involved in cooperative binding between the two domains. Cleavage at this site does not occur in apo and fully loaded states, but is significant in the partially loaded state. The induced susceptibility of E31 to cleavage is remarkably correlated to the induced protection from cleavage at E87, implicating that the observed changes are not local and possibly cooperative.

Furthermore, a structural homolog of the N-terminal domain of CaM is represented by troponin C (TnC). We have performed the structural alignment of TnC (PDB code 1avs: residues 15-87) and CaM (3cln: residues 5-77) which yields an RMSD of 1.0 Å. 70% of the aligned residues are identical, and 88% are homologous, making TnC a viable model for the N-domain of CaM. $Ca^{+2}$ loaded structure of TnC has been determined at 1.75 Å resolution [52]. Furthermore, single site E41A mutation in this protein and analysis by NMR indicates that there is direct coupling between binding of calcium to this particular EF-hand motif and the structural change induced [53]. We note that E41 was found to be strikingly unique in its control of TnC motions which is shown to single-handedly lock the large conformational change whereby several residues have to move by more than 15 Å. The structural alignment of CaM and TnC reveals that not only do E41 of TnC and E31 of CaM occupy analogous positions in terms of $Ca^{+2}$ ion coordination, they also both have the same overall EF-hand motif structure. We therefore assume that the critical role attributed to E41 in TnC is transferrable to E31 in CaM.

The similarity of these two residues is also corroborated by E31K mutations which do not lead to apparent binding affinity changes of $Ca^{+2}$ to CaM [54], as also occurs in the E41A mutation of TnC [53]. Conversely, E→K point mutations in the other three equivalent EF hand motif positions of CaM (E67K, E104K, and E140K) lead to the loss of $Ca^{+2}$ binding at one site [54]. Furthermore, E31K mutation has wild type activation on four different enzymes; smooth and skeletal muscle myosin light chain kinase (MLCK), adenylylcyclase, and plasma membrane $Ca^{2+}$-ATPase, while other mutants in the equivalent positions have poor activation [55]. Double mutants of these sites suggest a tight connection between loop I and loop IV, and this coupling is possibly mediated by the linker [56], since there is no NOE detected between N- and C-terminal lobe residues [57].

The connection between E31 location and the linker was later shown by a comparative MD study on $Ca^{+2}$ loaded CaM versus CaM where the $Ca^{+2}$ ion in EF-hand loop I is stripped from the structure. This study reveals that although the former is stable in its elongated form during the entire course of the simulation (12.7 ns), the lack of this particular $Ca^{+2}$ ion leads to structural collapse of the two domains at ca. 7.5 ns [46]. This change was observed to follow the loss of helicity in the linker region.

To further investigate the connection between loop I local structural changes and the linker, in figure 3a, we display the response profile of the perturbation that leads to the largest overlap between the experimental and predicted displacement profiles, $O^{31} = 0.72$. The direction of the applied force is displayed as a thick red arrow, and the response vector is shown by the thin green arrows. The overall bending of the two lobes towards each other is clear. We observe that the response is small in the first 1/3 portion of the linker, while it is magnified in the bottom 2/3, past R74 around which the linker has been noted to unwind even in early and much shorter (3 ns) simulations of CaM, possibly facilitating the reorientation of the two calcium binding domains [49]. In fact, more recent MD simulations of length 11.5 ns at physiological ionic strength revealed that the central helical region unwinds at ca. 3.5 ns, although the measured radius of gyration is consistent with the extended conformation throughout the simulation. The unwinding process involves the breaking of hydrogen bonds at residues 74-81 [47]. These authors observe rigid motions of the two domains around a single "hinge point" located here. Furthermore, pH titration experiments on CaM dimethylated with [$^{13}$C] formaldehyde imply that the pKa of Lys-75 is highly sensitive to the environmental changes such as peptide binding, indicating that the helical linker region unravels around this point [58]. Proteolysis of trypsin sensitive bonds lead to cleavage in Arg-74, Lys-75 and Lys-77 of the central helix which is not eliminated at high Ca$^{+2}$ concentrations, while at intermediate concentrations there is an order of magnitude increase in the rate of proteolysis indicating enhanced flexibility [59]. This behavior suggests that the linker may take on different roles depending on the solution conditions.

Perhaps equally important to simultaneously inducing bending and twisting motions by perturbing a single residue is the direction of the perturbation. All perturbations that give large overlaps with the targets fall along this line of perturbation within ±10º, making use of the less crowded region between this and helix A (residues 5-19). Although the region has low solvent accessibility due to the presence of side chains, this direction is nevertheless a convenient pathway for proton uptake/release. Huang and Cheung have studied in detail the effect of $H^+$ and $Ca^{+2}$ concentrations on activation of enzymes by calmodulin [60,61]. Their findings suggest that the addition of $Ca^{+2}$ exposes an amphipathic domain on CaM, whereas $H^+$ exposes a complimentary CaM binding domain on the target enzyme. The additional flux of $H^+$ might originate from CaM upon $Ca^{+2}$ binding, or from transient cell conditions or both. Their findings also suggest that such changes might occur via subtle pH changes in the range of $6.9 - 7.5$.

$Ca^{+2}$ ion in EF-hand loop I in CaM is modulated by three Asp and one Glu residue (figure 3b). In the absence of a direct electrostatic interaction between E31 and $Ca^{+2}$ ion, the side chain is expected to flip towards this gap, producing a local perturbation. A local scan of the possible isomeric states of the side chain of Glu31 yield only two possible conformations that will fit the gap; these alternative conformations are also shown on figure 3d as transparent traces.

Original backbone dynamics measurements made by $^{15}$N-NMR on $Ca^{+2}$-CaM indicated that the motion of the N- and C-terminal domains are independent [8]. Therein, a very high degree of mobility for the linker residues 78-81 is reported. These authors claim that their experiments support the idea the central linker acts merely as a flexible tether that keeps the two domains in close proximity. However, under different conditions, the two domains may well be communicating through the conformations assumed by the linker.

That helix A is stabilized upon calcium binding has been determined by frequency domain anisotropy measurements on unloaded and loaded CaM [62]. In the proposed model stemming from NMR analysis of a tetracysteine binding motif that has been engineered into helix A by site directed mutagenesis followed by fluorescent labeling, secondary structural changes in the linker orchestrate the release of helix A to allow for further $Ca^{+2}$ binding upon activation of the C-terminal domain. Results show that the large amplitude, nanosecond time scale motions occurring in this region are suppressed by $Ca^{+2}$ loading to the N-terminal domain. These results are also corroborated by binding kinetics studies on fluorescently labeled samples with various degrees of $Ca^{+2}$ loading [45]. Conversely, one may consider that helix A is destabilized once E31 side chain flips to release its grip on the $Ca^{+2}$ ion. We identify pH changes as a possible source for such local conformational changes.

**Degree of ionization calculations identify E31 as a proton uptake/release site at physiological pH range.** In recent years there has been accumulating evidence that conformations of proteins may be manipulated by their location in the cell; in particular pH variations in different cell compartments may be utilized for control. The pH may vary from as low as 4.7 in the lysozome to as high as 8.0 in the mitonchondria, with an average value of 7.2 which is also the value in the cytosol and the nucleus [63]. Adaptation to different pH values in various subcellular compartments [64] is thought to be directly related to protein stability [65] and pH of optimal binding affinity of interacting proteins [66]. Methods to determine how proteins adapt to cellular and sub-cellular pH are currently sought-after [67]. See ref. [68] for a review of how changes in pH under physiological conditions affect conformations of a variety of proteins and their functionality. In many cases differential changes on the order of $0.3 - 0.5$ units trigger the transformations.

To quantify if position 31 is particularly sensitive to subtle pH variations in the physiologically relevant range, we have calculated the degree of ionization of the charged amino acids by using the PHEMTO server [69,70]. There are 52 titratable groups in CaM, of which 36 are Asp or Glu, 13 are Lys or Arg, two are Tyr and one is a His. The variation in the degree of ionization as a function of pH is displayed in figure 4 for four types of charged amino acids. We find that, only two residues E31 and D122 have large variations in the range of physiologically relevant pH values. The upshift of E31 from the standard value of 4.4 is confirmed by all three other methods (PROPKA [71], H++ [72], and pKD [73]) as well as the experimental value reported from the structural homolog of calbindin and MCCE calculations [74] (see Table S1). We therefore propose that subtle changes in the pH of physiological environments may be utilized by the protonation/deprotonation of E31, whereby a local conformational change may be translated into the displacement profiles exemplified in figure 3, therefore leading to shifts in the conformational energy landscape.

Our findings indicate that at lower pH, E31 will be uncharged so that the side chain will not be stabilized by the $Ca^{+2}$ ion, and therefore will have a higher probability to occupy alternative conformations (Figure 3b). PRS shows that such a local conformational change propagates to the linker region and beyond to favor the compact forms. This finding is in agreement with the FRET experiments conducted at pH 5.0 versus 7.4 of $Ca^{+2}$-CaM where the distribution of distances between the fluorescently labeled donor (34)-acceptor (110) residues on either domain shifted significantly towards more compact conformations so that the extended conformation was almost entirely absent at reduced pH [13].

The pH effect has been noted as early as 1982, when the activation of MLCK by CaM was shown to occur in the pH range of 6.0 – 7.5 [75]. This is the range of pH where CaM is known to have the more rigid structure exposing the domain at the site on interaction. For the MLCK bound form, the catalytic activity exhibits a broad optimum from pH 6.5 to pH 9.0. This bound form is represented by the structure 2bbm in our set where PHEMTO calculations now find nine negatively-charged residues whose pKa values have upshifted to this range (D22, E31, D64,

E67, D95, E104, D131, D133 and E140); all except D64 are EF-hand loop $Ca^{+2}$ coordinating residues (see subsection Proteins under Materials and Methods).

Finally D122, which does not participate in EF-hand loops, but has a predicted pKa in the physiological range (gray curve in figure 4a) is found by PRS to be important for the conformational change between 3cln→1lin and 3cln→1prw (Table 3) with non-specific perturbation directions. It is plausible that this residue also acts as a local pH sensor for manipulating the conformations that favor closed form.

## DISCUSSION

We have studied the manipulation of the extended structure of $Ca^{+2}$ loaded CaM to seven different structures reported in literature. Due to the variety of functions performed by CaM, these represent different conformations that it may take on. Our main findings indicate the following: (i) Reduction of the CS to a few degrees of freedom conveniently describes the main features of the conformational changes of $Ca^{+2}$-CaM. These are represented via a simultaneous twist and bend motion of the two lobes with respect to each other (figure 2). (ii) For five of the seven structures, the conformational change occurs as a projection on the same vector set (Table 3), although the RMSD values may be large. The change is also independent of ligand size. (iii) This vector set, however is not simply described by a single underlying collective mode, but corresponds to some motion that seems to be stimulated by perturbing a particular residue (E31) in a particular direction (figure 3a). EndoGluC proteolysis [51] and a series of mutational studies [54,55,56,76] have uniquely identified E31 as a center influencing the dynamics of CaM. (iv) The perturbations of E31 induces coupled counter twisting and bending motions in the linker, and the bend is induced around residues found susceptible to dynamical changes via NMR [58] and proteolysis experiments [59]. (v) Independently, we find E31 to also be a unique residue (out of a total of 52 charged ones) whose ionization state is sensitive to subtle pH variations in the physiological range as corroborated with four different pKa calculation approaches. The transition between charged/uncharged states in E31 occurs in a narrow pH window of ca. 6.5 – 7.5. Combined with item (iv) above, E31 is thus implicated to be a center for conformation control via differential pH gradients.

These findings are in agreement with many experimental results obtained for $Ca^{+2}$-CaM in different environments, and further provide an explanation of the observations. We therefore propose a mode of functioning for CaM whereby it utilizes the pH differences in various compartments of the cell to perform different functions. At relatively high pH, the structure will be more compact, while at reduced pH values, gaining a proton at site E31 will possibly lead to a torsional jump changing the residue's side chain isomeric state. This will generate a shift in the conformational energy landscape, making compact conformational states more easily accessible.

The mechanism outlined above is not the sole means of conformational energy changes observed in CaM. In 1rfj (group 2) which rests on the reduced CS at a location quite close to the initial (extended) CaM structure, the conformational change is described by a simple bending; this is also the only liganded structure which is visited during the 120 ns long MD trajectory. For 1mux (group 3), on the other hand, either a single perturbation or a modal analysis will only partially describe the full conformational change, although the structure is located much closer on the CS to the initial form than the group 1 molecules (see figure 2 and Table 3). Both $\theta$ and $\varphi$ angles are only reduced by ca. 20°, in contrast to 1rfj where they are increased by ca. 20 and 40° respectively. This structure is never visited during the 120 ns long MD simulations and is possibly stabilized by the ligand WW7 (Table 1).

We emphasize that PRS is a technique that operates in the linear response regime, and therefore cannot account for further conformational changes, unless updated variance-covariance matrices in the shifted energy landscape are utilized. Nevertheless, the findings are not unique to CaM. Our previous study of FBP via PRS also indicated charged allosteric residues to coordinate ion release, again implicating a coupled electrostatic – mechanical effect [21]. In FBP, the remarkably high association constants on the order of $10^{17} - 10^{22}$ $M^{-1}$ suggests it is fairly easy for FBP to capture the ion, but also poses the question of how it is released once transported across the periplasm. Allosteric control using the different electrostatic environments in physiological conditions was put forth as a possible mechanism to overcome this so-called ferric binding dilemma [21]. It is remarkable that of the residues listed in Table 3, that produce the targeted conformational change via either direction specific or non-specific perturbations are predominantly charged (E6, K30, E31, R86, E119, D122, D123), although the electrostatics is not directly implemented in the PRS technique. That in both these proteins charged residues are revealed by solely analysis of the mechanical response of the protein makes PRS a promising method for studying conformational shifts near the free energy minima of proteins controlled by intracellular pH sensing.

In addition to NMR [15], x-ray [17], FRET [18], and single molecule force spectroscopy [4] methods, combined ion mobility/mass spectroscopy methods are also becoming attractive for investigating the effect of different environments on conformation distributions of proteins. A recent example is electro-spraying experiments on various CaM structures which indicate that the extended conformation is abundant at higher charge-states of the protein [14]. Developing simple and efficient methods such as PRS for investigating the relationship between modulated electrostatic environment of the protein and its mechanical response will therefore continue to be an important area of research, particularly as a unique approach to attack the problem of identifying adaptations of proteins to subcellular pH [67].

**MATERIALS and METHODS**

**Proteins.** CaM is a small acidic protein of 148 residues. In studies of $Ca^{+2}$-CaM, the extended conformation with a dumbbell shape containing two domains joined by an extended linker is customarily used (figure 1, boxed). Throughout the text we refer to the domains as N-terminal (or N-lobe) and C-terminal (or C-lobe). These include residues 5 – 68 and 92 – 147, respectively. Upon binding to the ligand, the domains move relative to each other and the flexible linker region changes conformation accordingly. All CaM structures studied in this work contain four $Ca^{2+}$ ions, two bound to each domain.

Thus, each domain has two helix-loop-helix $Ca^{2+}$-binding regions, referred to as EF-hand structure. This is a 12-residue-long highly conserved motif, whereby positions 1-3-5 are occupied by Asp or Asn residues which act as monodentate $Ca^{2+}$ ligands, and position 12 is occupied by a bidentate Asp or Glu. For the CaM structures studied here, the coordinating residues in each of the four EF-hands are as follows: loop I (D20-D22-D24-E31), loop II (D56-D58-N60-E67), loop III (D93-D95-N97-E104), loop IV (D129-D131-D133-E140).

Here we study the conformational change from this extended structure to a set of seven calmodulins; their three dimensional structures are also shown in figure 1. The bound ligands, the PDB codes of the target structures, the experimental resolution for those structures determined by x-ray methods, and the source organisms are listed in Table 1. The initial structure is the four-calcium-bound, open form of calmodulin, represented by the PDB structure 3cln.

**Perturbation Response Scanning.** The PRS method is based on the assumption that the ligand bound state of the protein is described by a perturbation of the Hamiltonian of the unbound state.

Under the linear response assumption, the shift in the coordinates is approximated by:

$$\Delta \mathbf{R}_1 = \left\langle \mathbf{R} \right\rangle_1 - \left\langle \mathbf{R} \right\rangle_0 \quad \frac{1}{k_B T} \left\langle \Delta \mathbf{R} \Delta \mathbf{R}^T \right\rangle_0 \Delta \mathbf{F} = \frac{1}{k_B T} \mathbf{C} \Delta \mathbf{F} \qquad (1)$$

where the subscripts 1 and 0 denote perturbed and unperturbed configurations of the protein, the angle brackets denote the ensemble average and superscript T denotes the transpose. $\Delta \mathbf{F}$ vector contains the components of the externally inserted force vectors on the selected residues. $\left\langle \Delta \mathbf{R} \Delta \mathbf{R}^T \right\rangle_0 = \mathbf{C}$ is the variance-covariance of the atomic fluctuations in the unperturbed state of the protein. In this study, we utilize MD simulations to represent the variance-covariance matrix which acts as the kernel between the inserted perturbations and the recorded displacements. Different derivations leading to the above equation may be found in references [21,43,77].

Our detailed PRS analysis is based on a systematic application of equation 1. We apply a random force to selected $C_\alpha$ atom(s) of the initial structure. We scan the protein using this strategy, consecutively perturbing each residue $i$ by applying the force $\Delta \mathbf{F}$ on $C_\alpha$ atom. Thus, $\Delta \mathbf{F}$ vector is formed in such a way that all the entries, except those corresponding to the residue being perturbed, are equal to zero. For a selected residue $i$, the random force $q_i$ is $(q_x\ q_y\ q_z)^i$ so that the external force vector is constructed as:

$$\Delta \mathbf{F}^T = \left\{ 0\ 0\ 0 \ldots \left( q_x\ q_y\ q_z \right)^i \ldots 0\ 0\ 0 \right\}_{1 \times 3N} \qquad (2)$$

We then compute the resulting changes $(\Delta \mathbf{R})^i$, as a result of the linear response of the protein, through equation 1. It is also possible to insert multiple perturbations to the protein by adding other triplets of non-zero terms to the $\Delta \mathbf{F}$ vector corresponding to the perturbation locations of interest.

The predictions of the average displacement of each residue, $j$, as a response of the system to inserted forces on residue $i$, $\Delta\mathbf{R}^i$ are compared with the experimental conformational changes $\Delta\mathbf{S}$ between the initial and the target PDB structure, e.g. the apo and the holo forms. For the $\Delta\mathbf{S}$ vector, the holo experimental structure is superimposed on the apo form, followed by the computation of the residue displacements in $x$-, $y$-, $z$-directions. The goodness of the prediction is quantified as overlap coefficient, $O^i$, for each perturbed residue by comparing the predicted and experimental displacements:

$$O^i = \frac{\Delta\mathbf{R}^i \cdot \Delta\mathbf{S}}{\left|(\Delta\mathbf{R}\cdot\Delta\mathbf{R})^i(\Delta\mathbf{S}\cdot\Delta\mathbf{S})\right|^{1/2}} \qquad (3)$$

This is the dot product of the two vectors, as a measure of the similarity of the direction in the predicted conformational change. In all calculations based on equation 1, unless otherwise specified, we report the averages over 500 independent runs where the forces are chosen randomly from a uniform distribution [21,22]. We also report the result from the best directionality. We note that all overlaps are calculated for the 143 residues present in the 3cln PDB file, since the locations of residues 1-4 and 148 are not reported. The locations of the four calcium ions are also included in these calculations. Thus, we have a total of 147 nodes perturbed in each scan.

**Molecular Dynamics Simulations.** We have performed MD simulation using 3cln as the initial structure of CaM which has 143 amino acids and 4 calcium ions. The system is solvated using the VMD 1.8.7 program with solvate plug-in version 1.2 [78]. The NAMD package is used to model the dynamics of the protein-water system [79]. The CharmM27 force field parameters are used for protein and water molecules [80]. Water molecules are described by the TIP3P model. The initial box has dimensions 96×60×62 Å containing ca. 35000 atoms neutralized by standard addition of ions. Long range electrostatic interactions are calculated by the particle mesh Ewald method [81]. The cutoff distance for non-bonded interactions is set to 12 Å with the switching function turned on at 10 Å. The RATTLE algorithm is used to fix bond lengths to their average values [82]. The system is first energy minimized with the conjugate gradients algorithm. During the MD simulation, periodic boundary conditions are used and the equations of motion are integrated using Verlet algorithm with a step size of 2 fs [81]. Temperature control is carried out by Langevin dynamics with a dampening coefficient of 5/ps. Pressure control is attained by a Langevin piston. Volumetric fluctuations are preset to be isotropic in the NPT runs. The system is run in the NPT ensemble at 1 atm and 310 K until volumetric fluctuations are stable to maintain the desired average pressure. In this case, this process requires a 1 ns long equilibration period. The run in the NPT ensemble is extended to a total of 120 ns. The coordinate sets are saved at 2 ps intervals for subsequent analysis, leading to $T = 60000$ snapshots.

The correlations between residue pairs derived from the MD trajectory are of particular interest. We consider the Cartesian coordinates of the $C_\alpha$ atoms and $Ca^{+2}$ ions recorded at each time step $t$ in the form of the $3N \times 1$ coordinate matrix, $\mathbf{R}(t)$ where $N$ is the number of nodes. This matrix does not contain the original coordinates from the MD trajectories, but instead is obtained by the best superposition of all the $T$ structures. A mean structure $<\mathbf{R}_i(t)>$ is defined as the average over these coordinate matrices. One can then write the positional deviations for each residue $i$ as a function of time and temperature, $\Delta \mathbf{R}_i(t) = \mathbf{R}_i(t) - <\mathbf{R}_i(t)>$. These are organized as the columns of the $3N \times T$ fluctuation trajectory matrix,

$$\Delta \mathbf{R} = \begin{bmatrix} \Delta\mathbf{R}_1(t_1) & \Delta\mathbf{R}_1(t_2) & \cdot & \Delta\mathbf{R}_1(t_T) \\ \Delta\mathbf{R}_2(t_1) & \Delta\mathbf{R}_2(t_2) & \cdot & \Delta\mathbf{R}_2(t_T) \\ \cdot & \cdot & \cdot & \cdot \\ & & & \\ \cdot & \cdot & \cdot & \cdot \\ \Delta\mathbf{R}_N(t_1) & \Delta\mathbf{R}_N(t_2) & \cdot & \Delta\mathbf{R}_N(t_T) \end{bmatrix} \qquad (4)$$

The $3N{\times}3N$ variance-covariance (or correlation) matrix $\mathbf{C} = \langle \Delta\mathbf{R}\Delta\mathbf{R}^{\mathrm{T}} \rangle_0$ is then calculated from the trajectory for subsequent use in equation 1.

**Modal analysis.** To determine the collective modes of motion, we decompose the variance-covariance as $\mathbf{C} = \mathbf{U}^{\mathrm{T}} \mathbf{\Lambda} \mathbf{U}$ where $\mathbf{\Lambda}$ is a diagonal matrix whose elements $\lambda_j$ are the eigenvalues of $\mathbf{C}$, and $\mathbf{U}$ is the orthonormal matrix whose columns $\mathbf{u}_j$ are the eigenvectors of $\mathbf{C}$. $\mathbf{C}$ has six zero eigenvalues corresponding to the purely translational and rotational motions. In modal analysis, for a given mode $j$, $\mathbf{u}_j$ are treated as displacement vectors. Inner products between the $3N$ elements of the $\mathbf{u}_j$ and $\Delta\mathbf{S}$ vectors are used to select the mode, $j$, which best describes the conformational change. To assess the quality of the modes obtained by $\mathbf{C}$, we use the overlap equation (equation 3) by replacing the displacement vector upon perturbation, $\Delta\mathbf{R}$ by the normal vector, $\mathbf{u}_j$.

To measure the convergence of the trajectories, we have studied the spectral properties of the variance-covariance (i) as a whole, (ii) as 12 portions of 10 ns each, (iii) as 6 portions of 20 ns each, (iv) as 4 portions of 30 ns each, and (v) as a whole for the last 90 ns portion. We check that the contributions of each eigenvector to the overall spectra converge in these trajectories. Results are displayed in figure S1. Since the first 20 ns portion of the trajectory displays a different quality than the rest (dashed curves in figure S1), we discard this portion as well as an additional 10 ns portion of the trajectory in the reported results. We also note that the first five eigenvectors account for more than 96 % of all the motions in each case.

**Overlap of target structures.** The overlap between the displacement vectors of two structures gives information on their similarity:

$$O^{jk} = \frac{\Delta \mathbf{S}^j \cdot \Delta \mathbf{S}^k}{\left|(\Delta \mathbf{S}^j \cdot \Delta \mathbf{S}^j)(\Delta \mathbf{S}^k \cdot \Delta \mathbf{S}^k)\right|^{1/2}} \tag{5}$$

Here, the superscripts $j$ and $k$ refer to different three dimensional structures, and $\Delta \mathbf{S}$ is the displacement vector between the initial structure and the target structure. $O^{jk}$ is a measure of the similarity of the directionality of the conformational change that occurs upon binding. While $O^{jk}$ = 1 represents perfect overlap of the directionality of the conformational change, the RMSD of two structures that have such an overlap value need not be zero. Two structures may be moving along the same vector, but if the amount of the move is varied, they would yield different RMSD values. For example, consider the simple hinge motion of three points in space, where the closing motion may have proceeded by either a small or a large amount. The RMSD between these two configurations would be large compared to their overlap, measured as the angle between the two lines of motion (also see figure 1 in ref. [33]). Thus, RMSD and overlap yield complementary information.

**p$K_a$ calculations.** For the p$K_a$ and degree of ionization calculations, we mainly used the PHEPS program implemented in the PHEMTO server [69,70]. The method is based on a self-consistent approach to calculating protein electrostatics. The intrinsic p$K_a$ value is defined as the modification of the p$K_a$ in the model compounds by the Born energy and the contributions from the partial charges of interacting atoms. Starting from a set of initial values, the electrostatic free energy is calculated iteratively until the p$K_a$ values converge. The effect of the bound Ca$^{+2}$ ions are also included in the calculations. We also perform p$K_a$ calculations using the PROPKA [71], H++ [72], and pKD [73] servers. PROPKA and pKD also rely on the accurate calculation of the shifts in free energies. The latter particularly focuses on a correct representation of hydrogen-bonding interactions, while the former is based on an improved description of the desolvation and the dielectric response of the protein. We note that while PROPKA version 3.0 has an improved representation of the titration behavior, it does not yet include ions explicitly in the calculations; we therefore used version 2.0 in the calculations. The H++ server uses a different approach than direct calculation of free energy changes, whereby the complicated titration curves are directly represented as a weighted sum of Henderson-Hasselbalch curves of decoupled quasi-sites. We report values for the settings of 0.15 M salinity, external dielectric of 80 and internal dielectric of 10. The latter is a suggested value for better prediction of solvent exposed residues, although we check that the trends are not affected by this choice. The calculated pKa values of CaM using PDB structure 3cln by these four methods are provided as supplementary Table S1. Where possible, experimentally measured values from literature are also included in that Table [83,84], along with reported calculations on the PDB structure 1cll using the MCCE method [74].

**ACKNOWLEDGEMENTS**

This work was supported by the Scientific and Technological Research Council of Turkey Project (grant 110T624). A.O.A. acknowledges Youssef Jameel Scholarship for her Ph.D. studies.

**Figure Captions**

**Figure 1.** Three dimensional structures of the proteins studied in the present work. The initial structure is placed in the center, and the targets are oriented in such a way that their C-terminal domains are best fitted. N-terminal domain is in green, C- terminal domain is in cyan, and the linker is in magenta. The $Ca^{2+}$ ions are shown as gray spheres, the bound ligand molecules are shown in gray surface representations.

**Figure 2. (a)** Schematic representations of the defined bending angles $\theta$, $\theta_N$, $\theta_C$, and the torsional angle $\varphi$. **(b)** $\theta/\varphi$ plot for the various target structures (filled circles) and the initial structure (empty circle). Joint probability distribution of the $\theta/\varphi$ pairs calculated from the 120 ns MD simulation (a total of 60000 conformations) is overlaid as a contour map. The initial structure resides in the most visited region of the conformational space during the simulation. Only the target structure 1rfj is visited within this time window.

**Figure 3.** (a) Best PRS prediction of the displacement vectors (green) belonging to the 3cln to 1lin conformational change, overlaid on the initial structure. The main motion is a bending of the two lobes as marked by the purple arrows accompanied by rotation of the linker, where motions are especially accentuated in the region of residues 75 – 90. (b) Coordination of the E31 related $Ca^{+2}$ ion. The side-chain ($\chi_1,\chi_2$) angles of E31 are in ($t,t$) conformation. Two alternate conformations of E31 which do not clash with any other heavy atoms in this conformation are also shown as transparent traces. These have ($t,g^-$) and ($g^-,g^+$) conformations for the ($\chi_1,\chi_2$) angle pair. The red thick arrow represents the best perturbation direction of E31 in both figures.

**Figure 4.** Degree of ionization as a function of pH for (a)Asp/Glu (36 residues) and (b)Lys/Arg (13 residues) amino acid types. E31 (dashed) and D122 (gray) are distinguished as those capable of changing ionization with subtle pH variations at physiological conditions. The standard pKa values for non-perturbed residues are 4.4 for Asp/Glu, 10.0 for Lys, 12.0 for Arg [68].

**Table 1.** target calmodulin structures studied in this work.

| PDB ID | source | sequence identity[1] | experimental resolution (Å) | ligand | reference |
|--------|--------|------------------|------------------------|--------|-----------|
| **1lin** | cow (bos taurus) | 99.3 | 2.0 | 4 trifluoperazine (TFP) groups, each having 23 heavy atoms | [3] |
| **1prw** | cow (bos taurus) | 99.3 | 1.7 | 1 acetyl group (3 heavy atoms) – represents the compact form of unliganded $Ca^{2+}$-CaM | [2] |
| **1qiw** | cow (bos taurus) | 98.6 | 2.3 | 2 diphenylpropyl-bis- butoxyphenyl ethyl-propylene-diamine (DPD) groups each having 28 heavy atoms | [85] |
| **2bbm** | drosophila melanogaster | 93.9 | NMR | myosin light chain kinase (26 residues) | [86] |
| **1cdl** | homo sapiens | 98.6 | 2.0 | protein kinase Type II alpha chain (19 residues) | [87] |
| **1rfj** | potato (solanum tuberosum) | 88.8 | 2.0 | 3 methyl-pentanediol (MPD) groups each having 8 heavy atoms | [88] |
| **1mux** | african frog (xenopus laevis) | 99.3 | NMR | 2 aminohexyl-chloro-naphthalenesulfonamide (WW7) groups, each having 22 heavy atoms | [89] |

[1] to the initial structure, PDB code 3cln from ratus ratus [16]

**Table 2.** RMSD between pairs of structures listed in Table 1. Lower diagonal: RMSD between overall structures; upper diagonal: RMSD between N-lobe (bold) and C-lobe (italic) domains only. Overlap between the experimental displacement vectors $\Delta \mathbf{S}$ are also displayed in parentheses.

| PDB ID | 3cln | 1lin | 1prw | 1qiw | 2bbm | 1cdl | 1rfj | 1mux |
|---|---|---|---|---|---|---|---|---|
| **3cln** | | **0.63** *0.70* | **2.3** *1.1* | **0.39** *0.95* | **1.9** *1.5* | **0.51** *0.76* | **0.63** *0.80* | **1.2** *1.3* |
| **1lin** | 15 | | **2.4** *0.70* | **0.68** *0.70* | **2.1** *1.2* | **0.51** *0.80* | **0.45** *0.91* | **1.1** *1.3* |
| **1prw** | 16 | 4.2 (0.96) | | **2.8** *0.70* | **3.3** *1.2* | **2.5** *1.0* | **2.1** *1.3* | **2.5** *1.5* |
| **1qiw** | 15 | 1.6 (0.99) | 3.6 (0.95) | | **1.8** *1.4* | **0.57** *0.90* | **0.68** *0.80* | **1.2** *1.1* |
| **2bbm** | 15 | 3.5 (0.94) | 5.2 (0.89) | 2.5 (0.96) | | **1.9** *1.4* | **2.0** *1.7* | **2.1** *1.4* |
| **1cdl** | 15 | 2.8 (0.97) | 4.6 (0.91) | 1.9 (0.99) | 2.2 (0.96) | | **0.48** *0.62* | **1.2** *1.1* |
| **1rfj** | 2.7 | 15 (0.09) | 16 (0.17) | 15 (0.09) | 15 (0.05) | 15 (0.03) | | **1.3** *1.1* |
| **1mux** | 6.4 | 15 (0.19) | 16 (0.17) | 14 (0.23) | 14 (0.26) | 14 (0.28) | 7.4 (0.20) | |

**Table 3.** Best overlap values obtained for proteins studied by PRS and modal analysis.*

| Protein Pair | Best PRS[a] | | Average PRS[a] | | Best mode | |
|---|---|---|---|---|---|---|
| | overlap | residue[b] | overlap[c] | residue | overlap | index[d] |
| 3cln/1lin | 0.72 | 30,31,69 | 0.54(17) | 122 | 0.48 | I |
| 3cln/1prw | 0.69 | 31,69 | 0.51(16) | 122 | 0.45 | I |
| 3cln/1qiw | 0.72 | 30,31,69 | 0.56(15) | 119 | 0.48 | I |
| 3cln/2bbm | 0.70 | 29,30,31,34 | 0.53(16) | 86 | 0.46 | I |
| 3cln/1cdl | 0.73 | 30,31,34 | 0.56(15) | 119 | 0.49 | I |
| 3cln/1rfj | 0.67 | 6 | 0.36(17) | 5 | 0.52 | II |
| 3cln/1mux | 0.43 | $Ca^{+2}$ in loop I | 0.23(9) | 123 | 0.30 | III |

* using the variance-covariance matrix obtained from the last 90 ns portion of the trajectory.

[a] results from 500 independent PRS runs.

[b] Those residues that lead to the largest overlaps ± 0.01 and appear at least 10 times with that largest overlap value are reported.

[c] Standard deviations shown in parentheses

[d] see text for mode shape classification and descriptions.

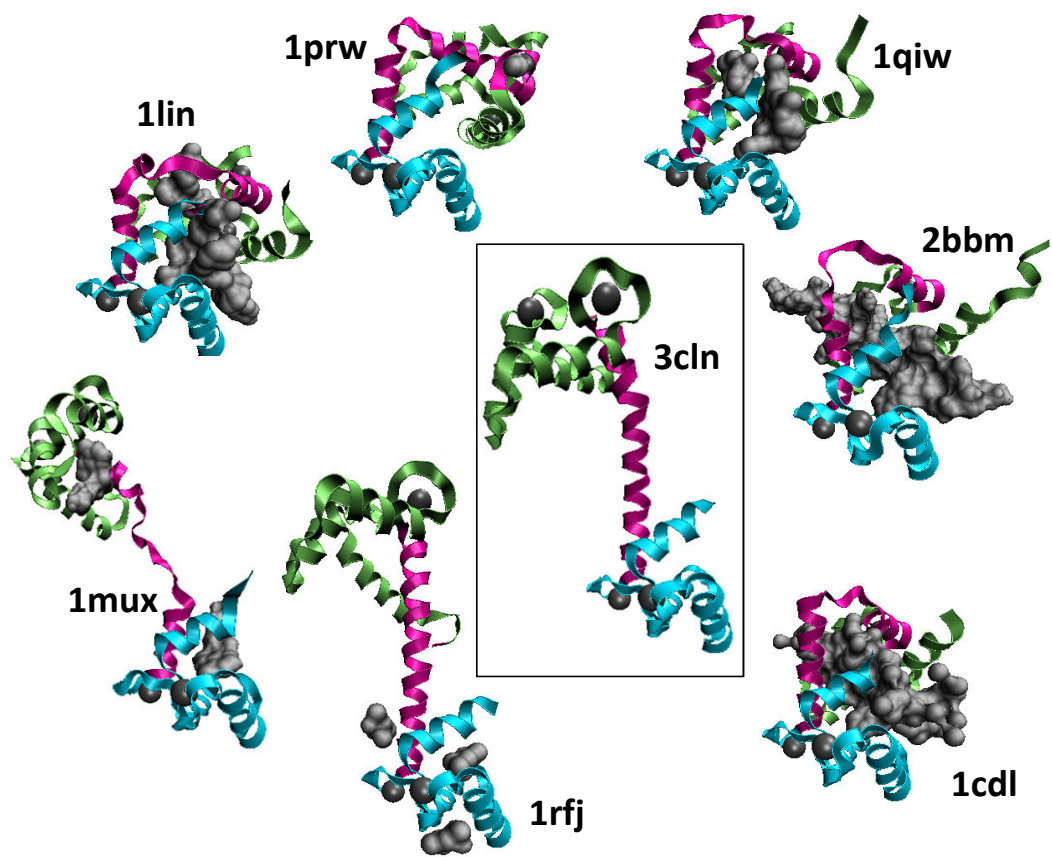

1lin 1prw 1qiw 2bbm 3cln 1mux 1rfj 1cdl

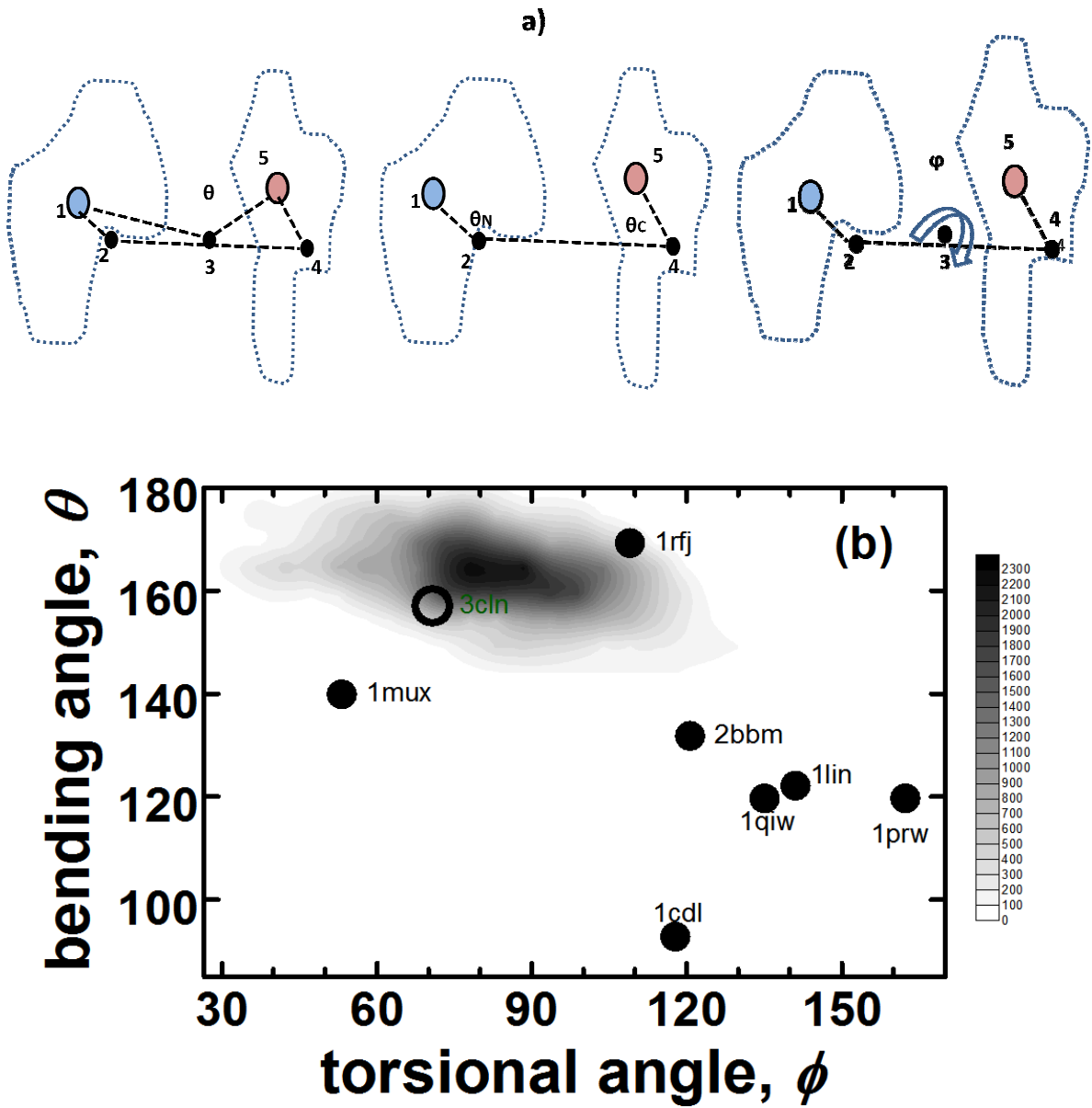

Figure 2

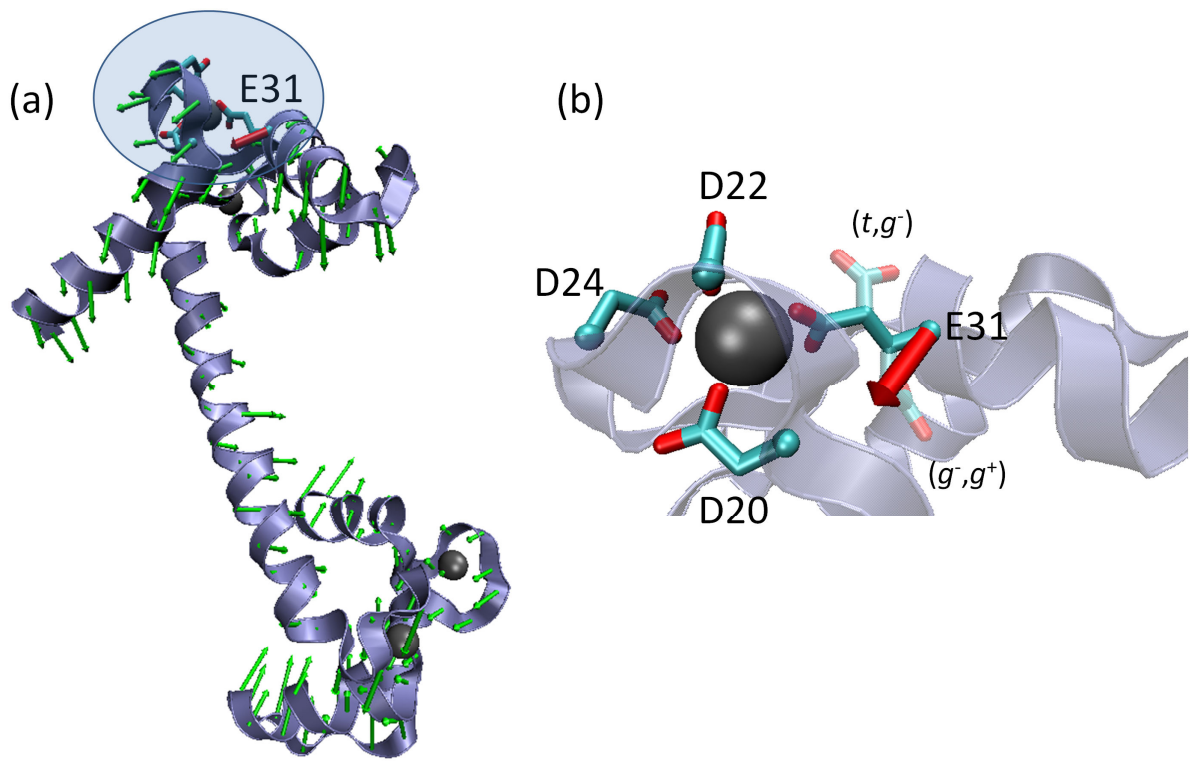

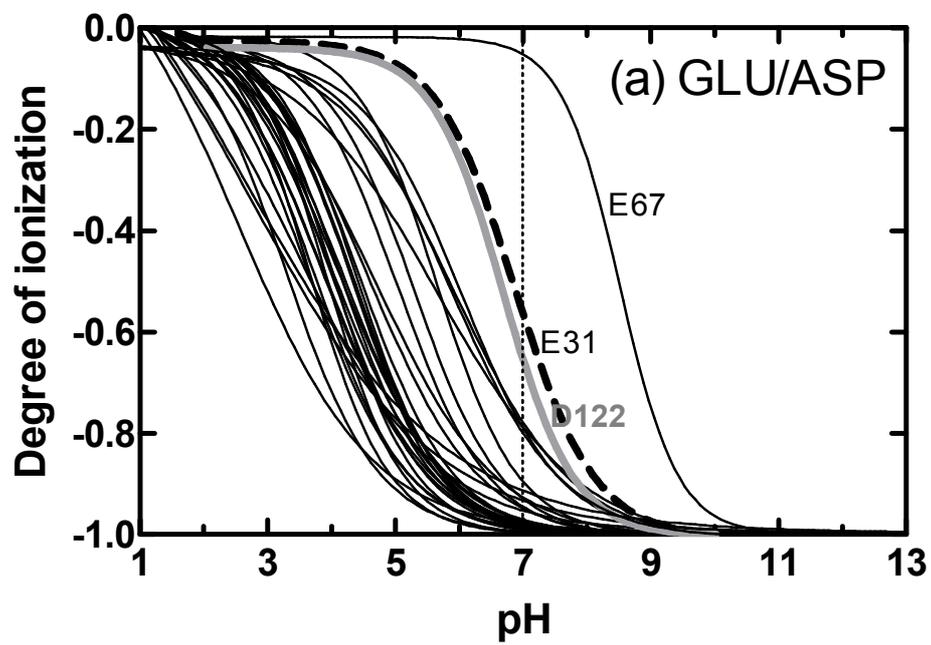

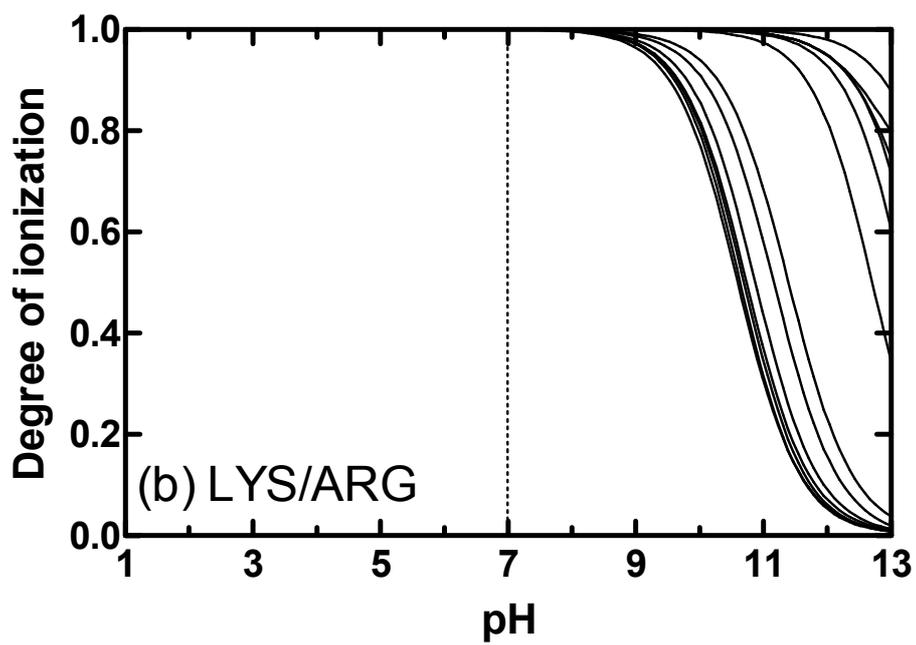